\documentclass[final]{siamltex}
\usepackage[lined,algonl,boxed]{algorithm2e}

\usepackage{hyperref}
\usepackage{mathdef}

\let\orightarrow\rightarrow
\let\omapsto\mapsto
\usepackage{breqn}
\let\rightarrow\orightarrow

\let\mapsto\omapsto
\usepackage{hyperbreqn}

\def\be{\begin{dmath*}}
\def\ee{\end{dmath*}}
\def\bel{\begin{dmath}}
\def\eel{\end{dmath}}
\def\bec{\begin{dmath*}[compact]}
\let\eec\ee
\def\belc{\begin{equation}}
\def\eelc{\end{equation}}

\def\bg{\begin{dgroup*}}
\def\eg{\end{dgroup*}}
\def\bgl{\begin{dgroup}}
\def\egl{\end{dgroup}}
\def\bs{\begin{dsuspend}}
\def\es{\end{dsuspend}}
\def\no{\hiderel}

\def\Re{\mathop{\rm Re}\nolimits}
\def\Im{\mathop{\rm Im}\nolimits}

\begin{document}

\title{Efficient Dealiased Convolutions without Padding\thanks{This work
was supported by the Natural Sciences and Engineering Research Council of
Canada.}}
\author{John C. Bowman\thanks{Department of Mathematical and Statistical
Sciences, University of Alberta, Edmonton, Alberta T6G 2G1, Canada.}
\and Malcolm Roberts$^\dagger$}

\maketitle

{\centerline {To appear in SIAM J. Sci. Comput., 2011}}

\begin{abstract}
Algorithms are developed for calculating dealiased linear convolution
sums without the expense of conventional zero-padding or phase-shift
techniques. For one-dimensional in-place convolutions, the memory
requirements are identical with the zero-padding technique, with the
important distinction that the additional work memory need not be
contiguous with the input data. This decoupling of data and work
arrays dramatically reduces the memory and computation time required
to evaluate higher-dimensional in-place convolutions. The technique
also allows one to dealias the higher-order convolutions that arise
from Fourier transforming cubic and higher powers.  Implicitly
dealiased convolutions can be built on top of state-of-the-art fast
Fourier transform libraries: vectorized multidimensional
implementations for the complex and centered Hermitian
(pseudospectral) cases have been implemented in the open-source
software {\tt FFTW++}.
\end{abstract} 

\begin{keywords} 
dealiasing, zero padding, convolution, ternary convolution, 
fast Fourier transform, bit reversal, implicit padding, pseudospectral
method, in-place transform, 2/3 padding rule 
\end{keywords}

\begin{AMS}
65R99, 65T50
\end{AMS}

\pagestyle{myheadings}

\bibliographystyle{siam}

\section{Introduction}
Discrete linear convolution sums based on the fast Fourier transform
(FFT) algorithm~\cite{Gauss1866,Cooley65} have become important tools
for image filtering, digital signal processing, and correlation
analysis. They are also widely used in periodic domains to solve
nonlinear partial differential equations, such as the Navier--Stokes
equations. In some of these applications, such as
direct numerical pseudospectral simulations of turbulent fluids,
memory usage is a critical limiting factor, and self-sorting in-place
multidimensional Fourier transforms~\cite{Temperton91} are typically used to
reduce the memory footprint of the required spectral convolutions.

It is important to remove aliases from FFT-based convolutions applied
to non\-periodic (wavenumber-space) data because they
assume cyclic input and produce cyclic output. Typically
the input data arrays are extended by padding them with enough zeros
so that wave beats of the positive frequencies cannot wrap around and
contaminate the negative frequencies. A cyclic convolution
$\sum_{p=0}^{N-1} f_p g_{k-p}$ is then performed using a
correspondingly larger Fourier transform size~$N$. If the cost of
computing a complex Fourier transform of size~$N$ is asymptotic to $K
N\log_2 N$ as \hbox{$N\goesto\infty$} (the lowest bound currently
achievable is $K=34/9$~\cite{Johnson07,Lundy07}), the asymptotic cost
of computing the convolution of two vectors of unpadded length~$m$ is
$6Km\log_2 m$ (using three Fourier transforms with $N=2m$).

Another important case in practice is the centered Hermitian~1D
convolution, dealiased by the {\it 2/3 padding rule} \cite{Orszag71}.
Since the computational cost of complex-to-real and real-to-complex Fourier
transforms of size~$N=3m$ is asymptotic to $\half K N\log_2 N$, the
FFT-based Hermitian convolution $\sum_{p=k-m+1}^{m-1} f_p g_{k-p}$
requires three transforms and hence $\fr{9}{2}Km\log_2 m$ operations.
Alternatively, phase shift dealiasing~\cite{Patterson71,Canuto06} can be used
to cancel out the aliasing errors between two convolutions with
different phase shifts. However, this second technique is rarely used in
practice since, in addition to doubling the memory requirements, it is
computationally more expensive, requiring $6K m\log_2 m$ operations for a 
centered Hermitian convolution.

An explicit application of the zero-padding technique involves the rather
obvious inefficiency of summing over a large number of data values that
are known {\it a~priori\/} to be zero.
However, it is worthwhile to consider the response
provided by Steven G. Johnson to a frequently asked question about the
possibility of avoiding this expense~\cite{fftwprune}:
\begin{quotation}\label{Johnson}
{\it
The most common case where people seem to want a pruned FFT is for
zero-padded convolutions, where roughly 50\% of your inputs are zero (to
get a linear convolution from an FFT-based cyclic convolution). Here, a
pruned FFT is hardly worth thinking about, at least in one dimension. In
higher dimensions, matters change (e.g.\ for a 3d zero-padded array about
1/8 of your inputs are non-zero, and one can fairly easily save a factor of
two or so simply by skipping 1d sub-transforms that are zero).
}
\end{quotation}
The reasoning behind the assertion that such one-dimensional pruned FFTs
are not worth thinking about is that if only $m$ of the $N$ inputs are nonzero,
the computational cost is reduced only from $N\log_2 N$ to $N\log_2 m$.
For example, if $m=N/2$, the savings is a minuscule $1/\log_2 N$.
Moreover, since a zero-padded Fourier transform of size~$N$ yields~$N$
(typically nonzero) output values, no storage savings appear
possible in one dimension. Nevertheless, in this work we demonstrate that
pruning the zero-padded elements of one-dimensional convolutions is still
worth thinking about, primarily because this provides useful building blocks
for constructing more efficient multidimensional convolutions.

The key observation is this: although the memory usage of our
implicitly padded 1D convolution is identical to that of a
conventional explicitly padded convolution, the additional temporary
memory need not be contiguous with the user data.  In a
multidimensional context, this external work buffer can be reused for
other low dimensional convolutions.  As a result, in $d$ dimensions an
implicitly dealiased convolution asymptotically uses $1/2^{d-1}$ of the
storage space needed for an explicitly padded convolution. When the
Fourier origin is centered in the domain, memory usage is reduced to
$(2/3)^{d-1}$ of the conventional amount.  If saving memory alone were
the goal, this reduction could also be achieved with explicit zero
padding by copying the data for the innermost convolution to an
external padded buffer, but such extra data communication would
degrade overall performance. The fact that our one-dimensional
convolution does not require this extra copying is the main feature
that was exploited to obtain simultaneous improvements in memory usage
and speed.

Nevertheless, the task of writing an efficient implicitly dealiased
one-dimensional convolution is onerous, particularly if one tries to
compete with a problem-dependent, architecture-adaptive
FFT algorithm (like that provided by the award-winning FFTW \cite{Frigo05}
library, which empirically predetermines a near optimal butterfly graph
at each subdivision). Effectively one wants to perform
the outer FFT subdivision manually, dropping the zero terms and
deferring the computation of the inner transforms to a standard library
routine. But this also restricts the set of available platform-dependent
algorithms that can be used at the highest level. Fortunately, several 
notable features of our algorithm help to offset this disadvantage. First, if
the goal is to  produce a convolution, bit reversal for the hand-optimized
outermost subdivision is unnecessary: the scrambled Fourier subtransforms of the
two input vectors can be multiplied together as they are produced
(perhaps while still accessible in the cache). Second, the implicit
method allows most of the subtransforms for an in-place convolution to
be optionally computed as out-of-place transforms, which typically execute
faster than in-place transforms (cf.~Figs.~1--6 of~\cite{Frigo05}) since they
require no extra (pre-, post-, or interlaced) bit-reversal stage.  These
savings help keep our one-dimensional in-place implicit convolution
competitive with an explicitly padded convolution based on the same highly
optimized library.

In Section~\ref{1d}, we develop an algorithm for Fourier transforming a
one-dimensional zero-padded vector without the need for explicit
padding. We show how this algorithm can be used to calculate implicitly
padded convolutions for both general and Hermitian inputs. 
We describe how implicit padding may be
applied to compute the discrete Fourier transform of an input
vector padded beyond an arbitrary fraction~$p/q$ of its length. 
Building on these one-dimensional algorithms, implicitly padded
convolutions are implemented for two- and three-dimensional input 
in Section~\ref{multid}. Finally, in Section~\ref{hyperconv}, we
show for both general and Hermitian data how implicit padding may be used to
dealias higher-order convolutions efficiently.

\section{Implicitly dealiased 1D convolutions}\label{1d}
In this section we describe the optimized 1D building blocks
that are used in subsequent sections to construct higher-dimensional
implicitly dealiased convolutions.

\subsection{Complex convolution}
The Fourier origin for standard convolutions is located at the first
(zero) index of the array.
Therefore, input data vectors of length~$m$ must be padded with zeros to
length $N\ge 2m-1$ to prevent modes with wavenumber $m-1$ from beating
together to contaminate mode~$N=0\mod N$. However, since FFT sizes with
small prime factors in practice yield the most efficient implementations,
it is normally desirable to extend the padding to $N=2m$.

In terms of the $N$th primitive root of unity, $\zeta_N\doteq \exp\(2 \pi
i/N\)$ (here $\doteq$ is used to emphasize a definition), the unnormalized backward
discrete Fourier transform of a complex vector
$\{U_k: k=0,\ldots,N\}$ may be written as
$$
u_j\doteq\sum_{k=0}^{N-1}\zeta_N^{jk} U_k,\qquad j=0,\ldots,N-1.
$$
The fast Fourier transform method exploits the properties that
$\zeta_N^r=\zeta_{N/r}$ and $\zeta_N^N=1$.

On taking $N=2m$ with $U_k=0$ for $k \ge m$, one can easily avoid
looping over the unphysical zero Fourier modes by decimating in wavenumber:
for $\ell=0,1,\ldots, m-1$:
\belc
u_{2\ell}
=\ds\sum_{k=0}^{m-1}\zeta_{2m}^{2\ell k} U_k
=\ds\sum_{k=0}^{m-1}\zeta_m^{\ell k} U_k,
\quad
u_{2\ell+1}
=\ds\sum_{k=0}^{m-1}\zeta_{2m}^{(2\ell+1) k} U_k
=\ds\sum_{k=0}^{m-1}\zeta_m^{\ell k} \zeta_{2m}^kU_k.
\label{cconv1backward} 
\eelc
This requires computing two subtransforms, each of size $m$,
for an overall computational scaling of order $2m\log_2 m=N\log_2 m$.

The odd and even terms of the convolution can then be computed separately
(without the need for a bit reversal stage), multiplied term by term, and
finally transformed again to Fourier space using the (scaled) forward transform
\ben
2mU_k&=&\sum_{j=0}^{2m-1}\zeta_{2m}^{-kj} u_j
=\sum_{\ell=0}^{m-1}\zeta_{2m}^{-k2\ell} u_{2\ell}
+\sum_{\ell=0}^{m-1}\zeta_{2m}^{-k(2\ell+1)} u_{2\ell+1}\endl
&=&\sum_{\ell=0}^{m-1}\zeta_m^{-k\ell} u_{2\ell}
+\zeta_{2m}^{-k}\sum_{\ell=0}^{m-1}\zeta_m^{-k\ell} u_{2\ell+1},
\qquad k\no=0,\ldots,m-1.\label{cconv1forward}
\een
The implicitly padded transforms described by~\Eqs{cconv1backward}
and~\hEq{cconv1forward} are implemented as
Procedure~{\tt\ref{fftpadBackward}} and~{\tt\ref{fftpadForward}}.
These algorithms are combined in Function~{\tt\ref{cconv}} to 
compute a dealiased convolution of unpadded length~$m$ using
two arrays of size~$m$ as input vectors instead of one array of size $2m$.
This seemingly trivial distinction is the key to the improved efficiency
and reduced storage requirements of the higher-dimensional implicit
convolutions described in Section~\ref{multid}.
Moreover, in Function~{\tt\ref{cconv}} we see that implicit padding allows
each of the six complex Fourier transforms of size $m$ to be done out of place.
In the listed pseudocode, an asterisk ($*$) denotes an element-by-element
(vector) multiply.

In principle, the stable trigonometric recursion described by
Buneman~\cite{Buneman87}, which requires two small precomputed tables, each
of size~$\log_2 N$, could be used to compute the required roots of unity
$\zeta_N^k$ that appear in \Eqs{cconv1backward}
and~\hEq{cconv1forward}.\footnote{We note that, in terms of the smallest
positive number $\e$
satisfying $1+\e > 1$ in a given machine representation, the singularity in
Buneman's scheme can be removed by replacing $\sec{\pi/2}$ with  $5/\e$,
$\sin 4\pi$ with $-\e/2$, and $\sin{2\pi}$ and $\sin{\pi}$ each with $-\e/10$.}
While Buneman's recursion has the same average accuracy as an FFT
itself~\cite{Tasche02}, on modern hardware a factorization method that does
not rely on successive table updates turns out to be more
efficient~\cite{Johnson09}, at the expense of somewhat higher memory usage.
We instead calculate the $\zeta_N^k$ factors with a single complex
multiply, using two short precomputed tables $H_a=\zeta_N^{as}$ and
$L_b=\zeta_N^b$, where $k=as+b$ with $s=\floor{\sqrt m}$,
\hbox{$a=0,1,\ldots,\ceil{m/s}-1$}, and $b=0,1,\ldots,s-1$. Since these
one-dimensional tables require only $\O(\sqrt{m})$ complex words of
storage and our focus is on higher-dimensional convolutions anyway, we do
not account for them in our storage estimates.

Referring to the computation times shown in Fig.~\ref{timing1c},
we see that the implicit padding algorithm described by \Eqs{cconv1backward} 
and~\hEq{cconv1forward}
can thus be implemented to be competitive with explicitly padded
convolutions. The error bars indicate the lower and upper {\it one-sided
standard deviations}
$$
\sigma_L=\sqrt{\fr{1}{\fr{n}{2}-1}}\sum_{i=1\atop{t_i < T}}^n (t_i-T)^2,
\qquad
\sigma_H=\sqrt{\fr{1}{\fr{n}{2}-1}}\sum_{i=1\atop{t_i > T}}^n (t_i-T)^2,
$$
where $T$ denotes the mean execution time of $n$ samples.
Both the FFTW-3.2.2 library and the convolution layer we built on
top of it were compiled with the Intel C/C++ 11.0 20081105 compiler, using
the optimization options {\tt -ansi-alias -malign-double -fp-model fast=2}
on a 64-bit 3GHz Intel E5450 Xenon processor with 6MB cache. Like the FFTW
library, our algorithm was vectorized for this architecture with
specialized single-instruction multiple-data (SIMD) code.

To compare the normalized error for the two methods, we considered the
input vectors $f_k=F e^{ik}$ and $g_k=G e^{ik}$ for $k=0,\ldots,m-1$,
with $F=\sqrt3+ i\sqrt 7$ and $G=\sqrt5+ i\sqrt {11}$.
The Fourier transforms of these vectors have nonzero components for all
transform sizes.
In Fig.~\ref{error1c} we compare the normalized~$L^2$ error 
$\sqrt{\sum_{k=0}^{m-1}\Abs{h_k-H_k}^2}/\sqrt{\sum_{k=0}^{m-1} \Abs{H_k}^2}$
for each of the computed solutions $h_k$ relative to the exact solution 
$H_k=\sum_{p=0}^k f_p g_{k-p}=FG (k+1) e^{ik}$.

\SetProcFnt{\textnormal}
\setlength{\algomargin}{0.6em}
\SetAlCapSkip{3pt}
\def\ProcNameFnt{\tt}

\def\fft{{\tt fft}}
\def\crfft{{\tt crfft}}
\def\rcfft{{\tt rcfft}}
\def\cconv{{\tt cconv}}
\def\conv{{\tt conv}}
\def\tconv{{\tt tconv}}
\def\build{{\tt build}}
\def\fftpadBackward{{\tt fftpadBackward}}
\def\fftpadForward{{\tt fftpadForward}}
\def\fftOpadBackward{{\tt fft0padBackward}}
\def\fftOpadForward{{\tt fft0padForward}}
\def\ffttpadBackward{{\tt ffttpadBackward}}
\def\ffttpadForward{{\tt ffttpadForward}}
\def\fftOtpadBackward{{\tt fft0tpadBackward}}
\def\fftOtpadForward{{\tt fft0tpadForward}}
\SetKwData{xf}{f}
\SetKwData{xu}{u}
\SetKwData{xFk}{F}
\SetKwData{xA}{A}
\SetKwData{xB}{B}
\SetKwData{xg}{g}
\SetKwData{xh}{h}
\SetKwData{xv}{v}
\SetKwData{xw}{w}
\SetKwData{xA}{A}
\SetKwData{xB}{B}
\SetKwData{xC}{C}
\SetKwData{xD}{D}
\SetKwData{xF}{F}
\SetKwData{xG}{G}
\SetKwData{xS}{S}
\SetKwData{xT}{T}
\SetKwData{xU}{U}
\SetKwData{xV}{V}
\SetKwData{xW}{W}

\begin{figure}[htbp]
\begin{minipage}{0.5\linewidth}
\begin{procedure}[H]
  \KwIn{vector \xf}
  \KwOut{vector \xf, vector \xu}
  \For{$k=0$ \KwTo $m-1$}{
    $\xu[k] \leftarrow \z_{2m}^k\xf[k]$\;
  }
  $\xf \leftarrow \fft\inv(\xf)$\;
  $\xu \leftarrow \fft\inv(\xu)$\;
  \caption{fftpadBackward({\sf f},{\sf u}) stores the scrambled
$2m$-padded backward Fourier transform values of a vector
{\sf f} of length $m$ in {\sf f} and an auxiliary vector~{\sf u} of length $m$.}\label{fftpadBackward}
\end{procedure}
\def\fftpadBackward{{\tt fftpadBackward}}
\begin{procedure}[H]
  \KwIn{vector \xf, vector \xu}
  \KwOut{vector \xf}
  $\xf \leftarrow \fft(\xf)$\;
  $\xu \leftarrow \fft(\xu)$\;
  \For{$k=0$ \KwTo $m-1$}{
    $\xf[k] \leftarrow \xf[k] + \z_{2m}^{-k}\xu[k]$\;
  }
  \Return f/(2m)\;
  \caption{fftpadForward({\sf f},{\sf u}) returns the
inverse of \fftpadBackward({\sf f},{\sf u}).}\label{fftpadForward} 
\end{procedure}
\end{minipage}
\begin{minipage}{0.49\linewidth}
\begin{function}[H]
  \KwIn{vector \xf, vector \xg}
  \KwOut{vector \xf}
  $\xu \leftarrow \fft\inv(\xf)$\;
  $\xv \leftarrow \fft\inv(\xg)$\;
  $\xu \leftarrow \xu * \xv$\;
  \For{$k=0$ \KwTo $m-1$}{
    $\xf[k] \leftarrow \z_{2m}^k\xf[k]$\;
    $\xg[k] \leftarrow \z_{2m}^k\xg[k]$\;
  }
  \medskip
  $\xv \leftarrow \fft\inv(\xf)$\;
  $\xf \leftarrow \fft\inv(\xg)$\;
  $\xv \leftarrow \xv * \xf$\;
  \medskip
  $\xf \leftarrow \fft(\xu)$\;
  $\xu \leftarrow \fft(\xv)$\;
  \medskip
  \For{$k=0$ \KwTo $m-1$}{
    $\xf[k] \leftarrow \xf[k] + \z_{2m}^{-k}\xu[k]$\;
  }
  \Return f/(2m)\;
\caption{cconv({\sf f},{\sf g},{\sf u},{\sf v}) computes
an in-place implicitly dealiased convolution of two complex vectors {\sf f}
and {\sf g} using two temporary vectors {\sf u} and {\sf v}, each of
length~$m$.}\label{cconv}
\end{function}
\end{minipage}
\end{figure}

\begin{figure}[htbp]
\begin{minipage}{0.49\linewidth}
\begin{center}
\includegraphics{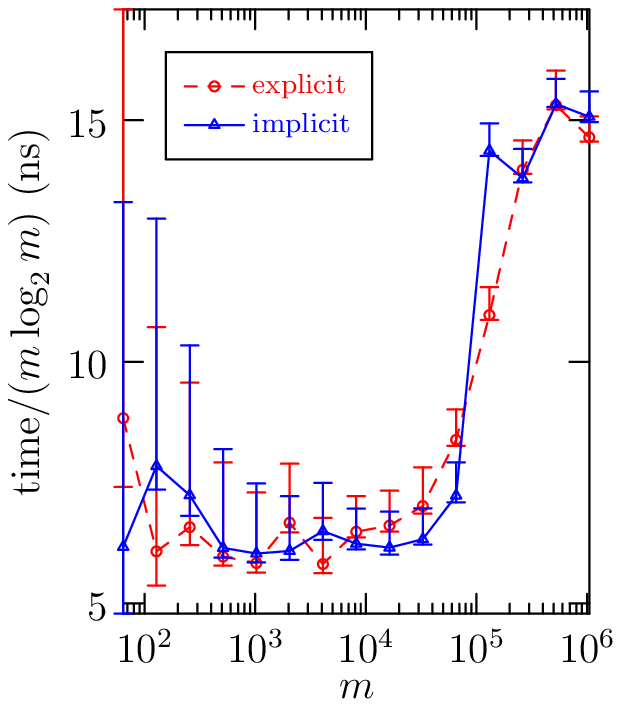}
\caption{Comparison of computation times for explicitly and implicitly
dealiased complex in-place 1D convolutions of length $m$. The storage
requirements of the two algorithms are identical.}
\phantomsection{}\label{timing1c}
\end{center}
\end{minipage}
\,
\begin{minipage}{0.49\linewidth}
\begin{center}
\includegraphics{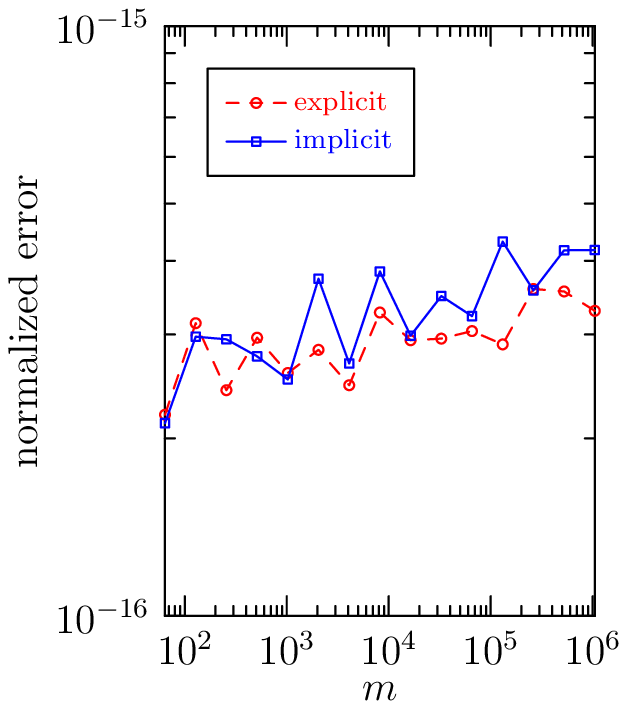}
\caption{Normalized $L^2$ error for explicitly and implicitly
dealiased complex in-place 1D convolutions of length $m$.}
\phantomsection{}\label{error1c}
\end{center}
\end{minipage}
\end{figure}

\subsection{Implicitly dealiased centered Fourier transform}\label{fft0}
A basic building block for constructing multidimensional centered
convolutions is an implicitly dealiased centered Fourier transform, where the
input data length is odd, say $2m-1$, with the Fourier origin at index $m-1$. 
Here, one needs to pad to $N\ge 3m-2$ to prevent
modes with wavenumber $m-1$ from beating together to contaminate
the mode with wavenumber $-m+1$. The ratio of the
number of physical to total modes, $(2m-1)/(3m-2)$, is asymptotic to $2/3$
for large $m$ \cite{Orszag71}.

For explicit padding, one usually chooses the padded vector length
$N$ to be a power of $2$, with $m=\floor{(n+2)/3}$.
However, for implicit
padding, it is advantageous to choose $m$ itself to be a power of $2$
since the algorithm reduces to computing FFTs of length $m$.
Moreover, it is convenient to pad implicitly slightly beyond $3m-2$, to $N=3m$,
as this allow the use of a radix $3$ subdivision at the outermost level, so
that only two of the three subtransforms of length $m$ need to be retained. 

Suppose then that $U_k=0$ for $k\ge m$.
On decomposing $j=(3\ell+r)\mod N$, where $r\in\{-1,0,1\}$, the 
substitution $k'=m+k$ allows us to write the backward transform as
\belc
u_{3\ell +r}\no=\sum_{k=-m+1}^{m-1}\z_m^{\ell k} \z_{3m}^{rk} U_k
=\sum_{k'=1}^{m-1}\z_m^{\ell k'} \z_{3m}^{r(k'-m)} U_{k'-m}
+\sum_{k=0}^{m-1}\z_m^{\ell k} \z_{3m}^{rk} U_k
=\sum_{k=0}^{m-1}\z_m^{\ell k} w_{k,r},\label{fft0backwardA}
\eelc
where
\bel
w_{k,r}\doteq
\cases{
U_0&if $k=0$,\cr
\z_{3m}^{rk}(U_k+\z_3^{-r}U_{k-m})&if $1\le k\le m-1$.\cr
}\label{fft0backwardB}
\eel
The forward transform is then
\bel
{3m}U_k=\sum_{r=-1}^{1}\zeta_{3m}^{-rk}\sum_{\ell=0}^{m-1}\zeta_m^{-\ell k} u_{3\ell+r},
\qquad k\no =-m+1,\ldots,m-1.\label{fft0forward}
\eel
The use of the remainder $r=-1$ instead of $r=2$ allows us to exploit
the optimization $\zeta_{3m}^{-k}=\conj{\zeta_{3m}^ k}$ in \Eqs{fft0backwardB}
and~\hEq{fft0forward}.
The number of complex multiplies needed to evaluate \Eq{fft0backwardB} 
for $r=\pm 1$ can be reduced by computing the intermediate complex quantities
\bg
\be
A_k\doteq \zeta_{3m}^k\(\Re U_k+\zeta_3^{-1} \Re U_{k-m}\),
\ee
\be
B_k\doteq i\zeta_{3m}^k\(\Im U_k+\zeta_3^{-1} \Im U_{k-m}\),
\ee
\eg
where $\zeta_3^{-1}=(-\half,-\fr{\sqrt{3}}{2})$, so that for $k > 0$,
$w_{k,1}=A_k+B_k$ and $w_{k,-1}=\conj{A_k-B_k}$.
The resulting transforms,
Procedures {\tt\ref{fft0padBackward}} and {\tt\ref{fft0padForward}},
each have an operation count asymptotic to $3Km\log_2 m$.
We were able to implement strided multivector versions of these algorithms
since they operate fully in place on their arguments, with no additional
storage requirements.

\begin{procedure}[htbp]
  \KwIn{vector \xf}
  \KwOut{vector \xf, vector \xu}
  $\xu[0] \leftarrow \xf[m-1]$\;
  \For{$k=1$ \KwTo $m-1$}{
    $\xA \leftarrow \zeta_{3m}^k\[\Re \xf[m-1+k]+\(-\half,-\fr{\sqrt{3}}{2}\)\Re \xf[k]\]$\;
    $\xB \leftarrow i\zeta_{3m}^k\[\Im \xf[m-1+k]+\(-\half,-\fr{\sqrt{3}}{2}\)\Im \xf[k]\]$\;
    $\xf[m-1+k] \leftarrow \xA+\xB$\;
    $\xu[k] \leftarrow \conj{\xA-\xB}$\;
    $\xf[0] \leftarrow \xf[k]$\;
    $\xf[k] \leftarrow \xf[k]+\xf[m-1+k]$\;
  }

  $\xf[0,\ldots,m-1] \leftarrow {\tt fft}\inv(\xf[0,\ldots,m-1])$\;
  $\xu[m] \leftarrow \xf[m-1]$\;
  $\xf[m-1] \leftarrow \xu[0]$\;
  $\xf[m-1,\ldots,2m-2] \leftarrow {\tt fft}\inv(\xf[m-1,\ldots,2m-2])$\;
  $\xu[0,\ldots,m-1] \leftarrow {\tt fft}\inv(\xu[0,\ldots,m-1])$\;
  \caption{fft0padBackward({\sf f},{\sf u}) stores the scrambled
$3m$-padded centered backward Fourier transform values of a vector {\sf f} of length
$2m-1$ in {\sf f} and an auxiliary vector~{\sf u} of length $m+1$.}\label{fft0padBackward}
\end{procedure}

\begin{procedure}[htbp]
  \KwIn{vector \xf, vector \xu}
  \KwOut{vector \xf}

  $\xf[m-1,\ldots,2m-2] \leftarrow {\tt fft}(\xf[m-1,\ldots,2m-2])$\;
  $\xu[m] \leftrightarrow \xf[m-1]$\;

  $\xf[0,\ldots,m-1] \leftarrow {\tt fft}(\xf[0,\ldots,m-1])$\;
  $\xu[0,\ldots,m-1] \leftarrow {\tt fft}(\xu[0,\ldots,m-1])$\;

  $\xu[m] \leftarrow \xf[0]+\xu[m]+\xu[0]$\;
  \For{$k=1$ \KwTo $m-1$}{
    $\xf[k-1]=\xf[k]+\(-\half,\fr{\sqrt{3}}{2}\)\zeta_{3m}^{-k}\xf[m-1+k]+\(-\half,-\fr{\sqrt{3}}{2}\)\zeta_{3m}^k\xu[k]$\;
    $\xf[m-1+k]=\xf[k]+\zeta_{3m}^{-k}\xf[m-1+k]+\zeta_{3m}^k\xu[k]$\;
  }
  $\xf[m-1] \leftarrow \xu[m]$\;
  \Return \xf/(3m)\;
  \caption{fft0padForward({\sf f},{\sf u}) returns the
inverse of \fftOpadBackward({\sf f},{\sf u}).}
\label{fft0padForward}
\end{procedure}

\subsection{Centered Hermitian convolution}

In this frequently encountered case (relevant to the pseudospectral
method), each input vector is the Fourier transform of real-valued data;
that is, it satisfies the {\it Hermitian symmetry} $U_{-k}=\conj{U_k}$.
While the physical data represented is of length $2m-1$, centered
about the Fourier origin, the redundant modes (corresponding to
negative wavenumbers) are not
included in the input vectors. The input vectors are thus of length $m$,
with the Fourier origin at index $0$. Just as in Section~\ref{fft0},
the unsymmetrized physical data needs to be padded with at least $m-1$ zeros.
Hermitian symmetry then requires us to pad the $m$ non-negative
wavenumbers with at least $c\doteq\floor{m/2}$ zeros.
The resulting $2/3$ padding ratio (for even $m$) turns out to work
particularly well for developing implicitly dealiased centered Hermitian
convolutions.
As in the centered case, we again choose the Fourier size to be $N=3m$.

Given that $U_k=0$ for $k\ge m$, the backward (complex-to-real) transform
appears as \Eq{fft0backwardA}, but now with
\bel
w_{k,r}\doteq
\cases{
U_0&if $k=0$,\cr
\z_{3m}^{rk}\(U_k+\z_3^{-r}\conj{U_{m-k}}\)&if $1\le k\le m-1$.\cr
}
\eel
We note that $w_{k,r}$ obeys the Hermitian symmetry 
$w_{k,r}=\conj{w_{m-k,r}}$, so that the Fourier transform
$\sum_{k=0}^{m-1}\z_m^{\ell k} w_{k,r}$ in \Eq{fft0backwardA} will indeed
be real valued. This allows us to build a backward implicitly dealiased
centered Hermitian transform using three complex-to-real Fourier transforms
of the first $c+1$ components of $w_{k,r}$ (one for each $r\in\{-1,0,1\}$). The
forward transform is given by
\bel
{3m}U_k
=\sum_{r=-1}^{1}\zeta_{3m}^{-rk}\sum_{\ell=0}^{m-1}\zeta_m^{-\ell k} u_{3\ell+r},
\qquad k\no =0,\ldots,m-1.\label{fft0forwardB}
\eel
Since $u_{3\ell+r}$ is real, a real-to-complex transform can be used to
compute the first $c+1$ frequencies of
$\sum_{\ell=0}^{m-1}\zeta_m^{-\ell k} u_{3\ell+r}$; the remaining $m-c-1$
frequencies needed in \Eq{fft0forwardB} are then computed using Hermitian
symmetry. 

Since there are two input vectors and
one output vector, the complete convolution requires a total of nine
Hermitian Fourier transforms of size $m$, for an overall computational
scaling of $\fr{9}{2}K m \log_2 m$ operations, in agreement with the
leading-order scaling of an explicitly padded centered Hermitian convolution.
For simplicity, we document here only the practically important
case $m=2c$; minor changes are required to implement the case $m=2c+1$.
We see in Function~{\tt\ref{conv}} that seven out of the nine Fourier
transforms can be performed out of place using the same amount of memory,
$2(\floor{N/2}+1)=6c+2$ words, as would be used to compute a centered Hermitian
convolution with explicit padding. 

To facilitate an in-place implementation of the
backward transform, we store the conjugate of the transformed values for
$r=1$ in reverse order in the upper half of the input vector,
using the identity (for real $u_j$)
$$
u_j=\conj{u_j}=\sum_{k=-c+1}^{c}\zeta_m^{-jk} \conj{U_k}
=\sum_{k'=m-1}^{0}\zeta_m^{j(k'-c)} \conj{U_{c-k'}}
=(-1)^j\sum_{k=0}^{m-1}\zeta_m^{jk} \conj{U_{c-k}}
$$
obtained with the substitution $k'=c-k$. One can omit the factors of
$(-1)^j$ here since they will cancel during the real term-by-term multiplication
of the two transformed input vectors.

\begin{figure}[htbp]
\begin{minipage}{0.51\linewidth}
\begin{procedure}[H]
  \KwIn{vector \xf}
  \KwOut{vector \xf, vector \xu}
  $\xu[0] \leftarrow \xf[0]$\;
  $\xFk \leftarrow \conj{f[2c-1]}$\;
  $\xf[2c-1] \leftarrow \xf[0]$\;
  \For{$k=1$ \KwTo $c-1$}{
    $\xA \leftarrow \zeta_{6c}^k 
\[\Re \xf[k]+\(-\half,\fr{\sqrt{3}}{2}\)\Re \xFk\]$\;
    $\xB \leftarrow -i\zeta_{6c}^k
\[\Im \xf[k]+\(-\half,\fr{\sqrt{3}}{2}\)\Im \xFk\]$\;
    $\xf[k] \leftarrow \xf[k]+\xFk$\;
    $\xu[k] \leftarrow \xA-\xB$\;
    $\xFk \leftarrow \conj{\xf[2c-1-k]}$\;
    $\xf[2c-1-k] \leftarrow \xA+\xB$\;
  }
  \caption{build({\sf f},{\sf u}) builds the FFT arrays
required for Function~{\tt\ref{conv}} from an unpadded vector {\sf f} of length
$2c$ into {\sf f} and an auxiliary vector~{\sf u} of length
$c+1$.}\label{build}
\end{procedure}
\end{minipage}
\begin{minipage}{0.49\linewidth}
\begin{center}
\includegraphics{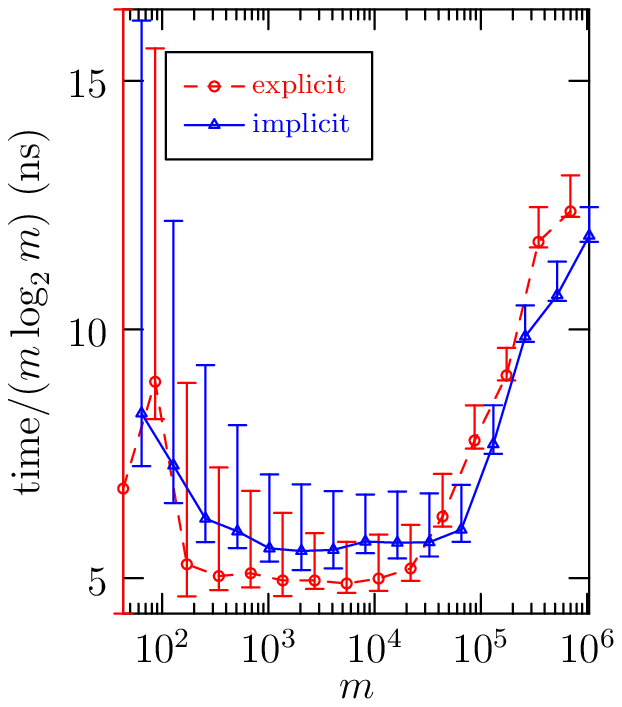}
\caption{Comparison of computation times for explicitly and implicitly
dealiased centered Hermitian in-place 1D convolutions of length $2m-1$.}
\phantomsection{}\label{timing1r}
\end{center}
\end{minipage}
\end{figure}

As seen in Fig.~\ref{timing1r}, the efficiency of the resulting implicitly
dealiased centered Hermitian convolution is comparable to an explicit
implementation. For each algorithm, we benchmark only those vector lengths
that yield optimal performance.

\begin{figure}[htbp]
\begin{center}
\includegraphics{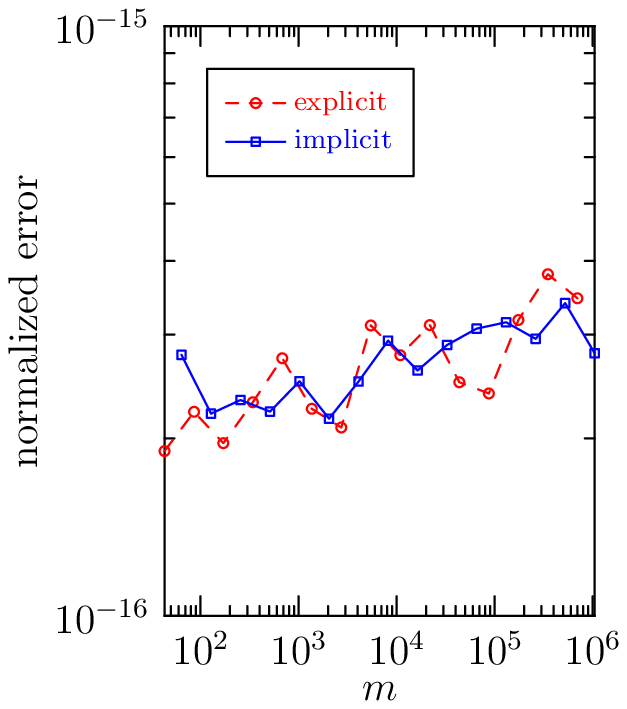}
\caption{Normalized $L^2$ error for explicitly and implicitly
dealiased centered Hermitian in-place 1D convolutions of length $m$.}
\phantomsection{}\label{error1r}
\end{center}
\end{figure}

To check the accuracy of our implementation, we used the test case
$f_k=F e^{ik}$ and $g_k=G e^{ik}$ for $k=0,\ldots,m-1$,
with $F=\sqrt 3$ and $G=\sqrt 5$, noting that Hermitian symmetry
requires that $F$ and $G$ be real. The exact solution is 
$H_k=FG\sum_{p=k-m+1}^{m-1} e^{ip}e^{i(k-p)}=FG(2m-1-k)e^{ik}$.
The normalized $L^2$ errors for implicit and explicit padding are compared
in Fig.~\ref{error1r}.

\begin{function}[htbp]
  \KwIn{vector \xf, vector \xg}
  \KwOut{vector \xf}
  $\xF \leftarrow \xf[c]$\;
  \build(\xf,\xu)\;
  $\xC \leftarrow \xf[c]$\;
  $\xf[c] \leftarrow 2\Re \xF$\;
  $\xu[c] \leftarrow \Re \xF+\sqrt{3}\Im \xF$\;
  \medskip
  $\xG \leftarrow \xg[c]$\;
  \build(\xg,\xv)\;
  $\xD \leftarrow \xg[c]$\;
  $\xg[c] \leftarrow 2\Re \xG$\;
  $\xv[c] \leftarrow \Re \xG+\sqrt{3}\Im \xG$\;
  \medskip
  $\xu \leftarrow \crfft(\xu)$\;
  $\xv \leftarrow \crfft(\xv)$\;
  $\xv \leftarrow \xv * \xu$\;
  $\xu \leftarrow \rcfft(\xv)$\;
  \medskip
  $\xv \leftarrow \crfft(\xf[0,\ldots,c])$\;
  $\xf[0,\ldots,c] \leftarrow \crfft(\xg[0,\ldots,c])$\;
  $\xv \leftarrow \xv * \xf[0,\ldots,c]$\;
  $\xf[0,\ldots,c] \leftarrow \rcfft(\xv)$\;
  \medskip
  $\xS\leftarrow \xf[c-1]$\;
  $\xT\leftarrow \xf[c]$\;
  $\xf[c-1]=\Re \xF-\sqrt{3}\Im \xF$\;
  $\xf[c]=\xC$\;
  $\xg[c-1]=\Re \xG-\sqrt{3}\Im \xG$\;
  $\xg[c]=\xD$\;
  \medskip
  $\xv \leftarrow \crfft(\xg[c-1,\ldots,2c-1])$\;
  $\xg[c-1,\ldots,2c-1] \leftarrow \crfft(\xf[c-1,\ldots,2c-1])$\;
  $\xg[c-1,\ldots,2c-1] \leftarrow \xg[c-1,\ldots,2c-1] * \xv$\;
  $\xv\leftarrow \rcfft(\xg[c-1,\ldots,2c-1])$\;
  \medskip

  \For{$k=1$ \KwTo $c-2$}{
    $\xf[k]=\xf[k]+\zeta_{6c}^{-k}\xv[k]+\zeta_{6c}^k\xu[k]$\;
    $\xf[2c-k]=\conj{\xf[k]}+\(-\half,-\fr{\sqrt{3}}{2}\)\zeta_{6c}^k\conj{\xv[k]}+\(-\half,\fr{\sqrt{3}}{2}\)\zeta_{6c}^{-k}\conj{\xu[k]}$\;
  }

  $\xf[c-1]=S+\zeta_{6c}^{1-c}\xv[c-1]+\zeta_{6c}^{c-1}\xu[c-1]$\;
  $\xf[c]=T-\(-\half,\fr{\sqrt{3}}{2}\)\xv[c]-\(-\half,-\fr{\sqrt{3}}{2}\)\xu[c]$\;
  \If{$c > 1$}{
    $\xf[c+1]=\conj{S}+\(-\half,-\fr{\sqrt{3}}{2}\)\zeta_{6c}^{c-1}\conj{\xv[c-1]}+\(-\half,\fr{\sqrt{3}}{2}\)\zeta_{6c}^{1-c}\conj{\xu[c-1]}$\;
  }

  \Return f/(6c)\;
\caption{conv({\sf f},{\sf g},{\sf u},{\sf v}) uses Procedure~\build\ to compute
an in-place implicitly dealiased convolution of centered Hermitian vectors
{\sf f} and {\sf g} of length~$2c$ using temporary vectors {\sf u} and
{\sf v} of length $c+1$.}\label{conv}
\end{function}

\subsection{General padding}\label{pq}
Implicit padding corresponding to an arbitrary~$p/q$ rule is also
possible. Suppose that $pm$ data modes are zero padded to $N=qm$, where $p$
and $q$ are relatively prime. One decomposes $j=q\ell+r$, where
$\ell=0,\dots,m-1$ and $r=0,\dots,q-1$.
Similarly, one expresses $k=t m+s$, where $t=0,\dots,p-1$ and $s=0,\dots, m-1$:
$$
u_{q\ell+r}=\sum_{k=0}^{pm-1}\z_{qm}^{(q\ell+r)k} U_k
=\sum_{s=0}^{m-1}\sum_{t=0}^{p-1} \z_m^{\ell (t m+s)}\z_{qm}^{r(t m+s)}
U_{t m+s}
=\sum_{s=0}^{m-1}\z_m^{\ell s}\sum_{t=0}^{p-1}\z_{qm}^{r(t m+s)} U_{t m+s}.
$$
Since there are $q$ choices of $r$, the problem reduces
to computing $q$ Fourier transforms of length~$m$, which requires
$K q m\log_2 m=K N\log_2 (N/q)$ operations. 
Likewise, the forward implicit transform
$$
{qm}U_k
=\sum_{\ell=0}^{m-1}\sum_{r=0}^{q-1} \zeta_{qm}^{-(q\ell+r)k} u_{q\ell+r}
=\sum_{r=0}^{q-1}\zeta_{qm}^{-rk}\sum_{\ell=0}^{m-1}\zeta_m^{-k\ell} u_{q\ell+r},
\qquad k\no =0,\ldots,pm-1
$$
also requires $q$ Fourier transforms of length $m$. Again, the computational
savings for a one-dimensional transform is marginal.

\section{Higher-dimensional convolutions}\label{multid}
The algorithms developed in Section~\ref{1d} can be used as building
blocks to construct efficient implicitly padded higher-dimensional convolutions.
\subsection{Complex 2D convolution}
The implicitly padded 2D convolution shown in Function {\tt\ref{cconv2}} 
is designed for data stored with a stride of one in the $y$
direction. Efficient multivector versions of Procedures {\tt\ref{fft0padBackward}}
and {\tt\ref{fft0padForward}} are used for the transform in the $x$ direction;
this allows a single $\zeta_{3m}^k$ factor to be applied to a consecutive
column of data at constant $x$. In principle, one could also develop a
multivector version of Function~{\tt\ref{cconv}} to perform simultaneous
convolutions in the $y$ direction, but our timing tests indicate that this
would only slightly enhance the overall performance (since the data for constant
$y$ is not stored consecutively in memory) and would prevent the 1D convolution
work arrays from being reused for the~$y$ convolution at each fixed $x$. The
memory savings in our method comes precisely from this reuse of temporary
storage, which in turn requires that the~$y$ convolutions be computed
serially.

\begin{figure}[htbp]
\begin{center}
\includegraphics{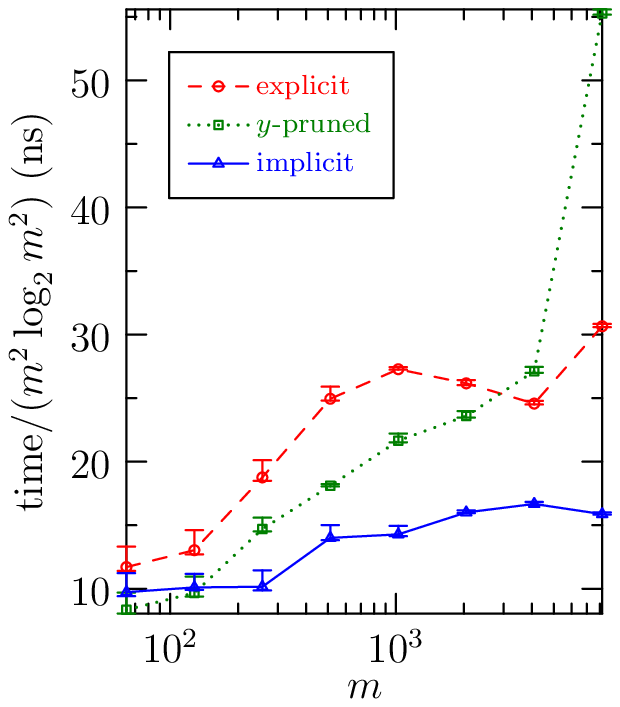}
\caption{Comparison of computation times for explicitly and implicitly
dealiased complex in-place 2D convolutions of size $m^2$.}
\phantomsection{}\label{timing2c}
\end{center}
\end{figure}

As shown in Fig.~\ref{timing2c}, the resulting implicit
2D algorithm dramatically outperforms the explicit version:
a $1024^2$ implicit complex convolution is $1.91$ times faster.

The third sentence of the quote from Steven G. Johnson on
page~\ref{Johnson} suggests a sensible optimization for
explicitly padded 2D convolutions: one can omit for $m \le y < 2m$
the backward and forward Fourier transforms in the $x$ direction.
However potential data locality optimizations may be lost
when a 2D convolution is expressed directly in terms of 1D transforms: as
observed in Fig.~\ref{timing2c}, while a $1024^2$ $y$-pruned explicit
convolution is $1.26$ times faster than a conventional explicit
implementation, the pruned method becomes $1.80$ times {\it slower} for the
$8192^2$ case. Our implicitly dealiased convolution is also subject
to these same optimization losses, but the savings due to implicit padding,
out-of-place subtransforms, the neglect of high-level bit reversal, and the
immediate convolution of constant $x$ columns (while still possibly in
cache) outweigh these losses.

Because the same temporary arrays $u$ and $v$ are used for each column
of the convolution, the memory requirement is $4m_xm_y+2m_y$ complex
words, far less than the $8m_xm_y$ complex words needed for an
explicitly padded convolution.

\subsection{Centered Hermitian 2D convolution}

In two dimensions, the Fourier-centered Hermitian symmetry appears as
$U_{-k,-\ell}=\conj{U_{k,\ell}}$. 
This symmetry is exploited in the centered Hermitian convolution
algorithm described in Function~{\tt\ref{conv2}}. As shown in
Fig.~\ref{timing2r}, implicit padding again yields a dramatic improvement
in speed.

When $m_y$ is even, the memory usage for an implicitly dealiased
$(2m_x-1)\times (2m_y-1)$ centered Hermitian convolution is
$2(2m_x-1)m_y+2(m_x+1)m_y+2(m_y/2+1)=6m_xm_y+m_y+2$ complex words, compared
with a minimum of $2(3m_x-2)(3m_y/2)=9m_xm_y-6m_y$ complex words required
for an explicitly dealiased convolution.

\begin{figure}[htbp]
\begin{minipage}{0.5\linewidth}
\begin{function}[H]
  \KwIn{matrix \xf, matrix \xg}
  \KwOut{matrix \xf}
  \For{$j=0$ \KwTo $m_y-1$}{
    $\fftpadBackward(\xf^T[j],\xU^T[j])$\;
    $\fftpadBackward(\xg^T[j],\xV^T[j])$\;
  }
  \For{$i=0$ \KwTo $m_x-1$}{
    $\cconv(\xf[i],\xg[i],\xu,\xv)$\;
    $\cconv(\xU[i],\xV[i],\xu,\xv)$\;
  }
  \For{$j=0$ \KwTo $m_y-1$}{
    $\fftpadForward(\xf^T[j],\xU^T[j])$\;
  }
  \Return \xf\;
\caption{cconv2({\sf f},{\sf g},{\sf u},{\sf v},{\sf U},{\sf V}) 
returns an in-place implicitly dealiased convolution of
\hbox{$m_x\times m_y$} matrices {\sf f} and {\sf g} using temporary 
\hbox{$m_x\times m_y$}
matrices {\sf U} and {\sf V} and temporary vectors {\sf u} and {\sf v} of
length $m_y$.}\label{cconv2}
\end{function}
\end{minipage}
\begin{minipage}{0.5\linewidth}
\begin{function}[H]
  \KwIn{matrix \xf, matrix \xg}
  \KwOut{matrix \xf}
  \For{$j=0$ \KwTo $m_y-1$}{
    $\fftOpadBackward(\xf^T[j],\xU^T[j])$\;
    $\fftOpadBackward(\xg^T[j],\xV^T[j])$\;
  }
  \For{$i=0$ \KwTo $2m_x-2$}{
    $\conv(\xf[i],\xg[i],\xu,\xv)$\;
  }
  \For{$i=0$ \KwTo $m_x$}{
    $\conv(\xU[i],\xV[i],\xu,\xv)$\;
  }
  \For{$j=0$ \KwTo $m_y-1$}{
    $\fftOpadForward(\xf^T[j],\xU^T[j])$\;
  }
  \Return \xf\;
\caption{conv2({\sf f},{\sf g},{\sf u},{\sf v},{\sf U},{\sf V}) 
returns an in-place implicitly dealiased centered Hermitian convolution of
$(2m_x-1)\times m_y$ matrices {\sf f} and {\sf g} using temporary
$(m_x+1)\times m_y$ matrices {\sf U} and~{\sf V} and vectors {\sf u} and
{\sf v} of length~$m_y$.
}\label{conv2}
\end{function}
\end{minipage}
\end{figure}

\begin{figure}[htbp]
\begin{minipage}{0.49\linewidth}
\begin{center}
\includegraphics{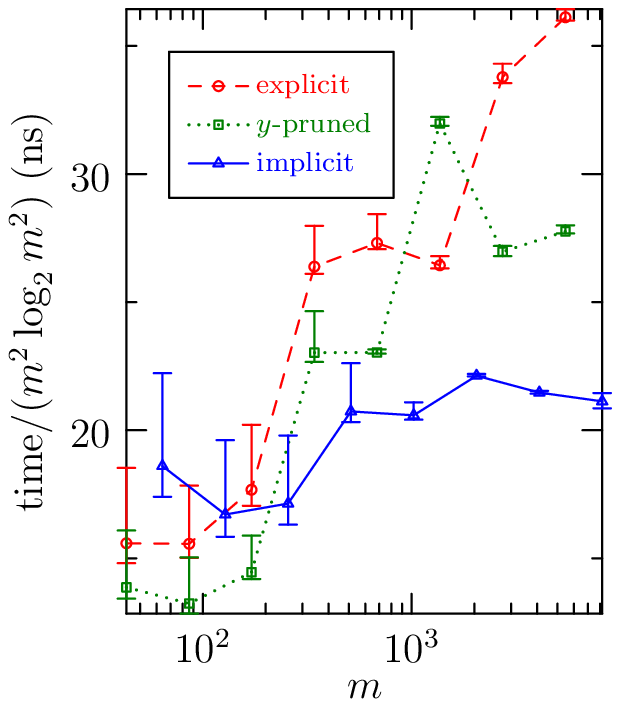}
\caption{Comparison of computation times for explicitly and implicitly
dealiased centered Hermitian in-place 2D convolutions of size \hbox{$(2m-1)\times m$}.}
\phantomsection{}\label{timing2r}
\end{center}
\end{minipage}
\,
\begin{minipage}{0.49\linewidth}
\begin{center}
\includegraphics{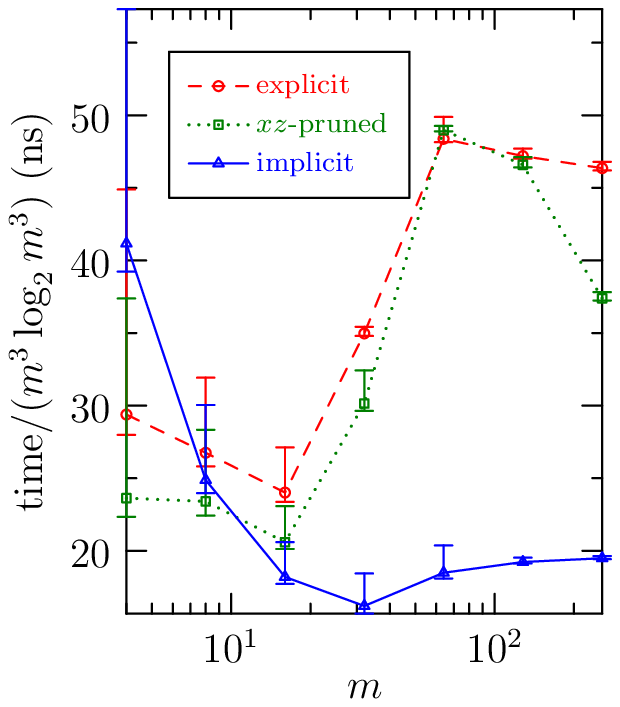}
\caption{Comparison of computation times for explicitly and implicitly
dealiased complex in-place 3D convolutions of size $m^3$.}
\phantomsection{}\label{timing3c}
\end{center}
\end{minipage}
\end{figure}

\subsection{2D pseudospectral application}
In our implementation, we allow the
convolution inputs to be arrays of vectors, $f_i$ and~$g_i$ ($i=1,\ldots,M$),
interpreting in Functions~{\tt\ref{cconv}},~{\tt\ref{conv}},
and~{\tt\ref{tconv}}, the product $f * g$ as the element-by-element
dot product $\sum_i f_i * g_i$.
Convolving $M$ input data blocks simultaneously like this enables, for example,
the nonlinear term of the 2D incompressible Euler equation
to be computed using five Fourier transforms (instead of three for
each of the $M=2$ input pairs). Specifically, the advective term
$-\vu\dot\grad\w=-(\zhat\cross\grad \del^{-2}\w)\dot \grad \w$,
which appears in Fourier space as
$$
\sum_{\vp} \frac{p_xk_y - p_y k_x}{|\vk-\vp|^2}\w_{\vp}\w_{\vk-\vp},
$$
can be computed efficiently with the call
${\tt conv2}(i k_x\w,i k_y\w,i k_y \w/k^2,-ik_x\w/k^2,{\sf u},{\sf v})$,
where {\sf u} and {\sf v} are work arrays.

\subsection{Complex and centered Hermitian 3D convolutions}

The decoupling of the 2D work arrays in Function~{\tt\ref{cconv2}}
facilitates the construction of an efficient 3D implicit complex convolution,
as described in Function~{\tt\ref{cconv3}}. As shown in
Fig.~\ref{timing3c}, an implicit
$256^3$ convolution is $2.38$ times faster than the explicit version, while
an $xz$-pruned version is only $1.24$ times faster. The memory usage of an
implicitly padded 3D $m_x\times m_y\times m_z$ convolution is
$4m_xm_ym_z+2m_y m_z+2m_z$ complex words, far less than the
$16m_xm_ym_z$ complex words required by an explicit 
convolution based on power-of-two transforms.

A $(2m_x-1)\times (2m_y-1)\times (2m_1-1)$ implicit centered Hermitian 3D
convolution was also implemented in an analogous manner. It required
$$
6m_x(2m_y-1)m_z+2(m_y+1)m_z+2(m_z/2+1)=12m_xm_ym_z-6m_xm_z+2m_ym_z+m_z+2
$$
complex words, in comparison with the usual requirement of $27m_xm_ym_z$
complex words for explicit padding with power-of-two transforms.

\begin{figure}[htbp]
\begin{minipage}{0.55\linewidth}
\begin{function}[H]
  \KwIn{matrix \xf, matrix \xg}
  \KwOut{matrix \xf}
  \For{$j=0$ \KwTo $m_y-1$}{
    \For{$k=0$ \KwTo $m_z-1$}{
      $\fftpadBackward(\xf^T[k][j],\xU^T[k][j])$\;
      $\fftpadBackward(\xg^T[k][j],\xV^T[k][j])$\;
    }
  }
  \For{$i=0$ \KwTo $m_x-1$}{
    $\cconv2(\xf[i],\xg[i],\xu_1,\xv_1,\xu_2,\xv_2)$\;
    $\cconv2(\xU[i],\xV[i],\xu_1,\xv_1,\xu_2,\xv_2)$\;
  }
  \For{$j=0$ \KwTo $m_y-1$}{
    \For{$k=0$ \KwTo $m_z-1$}{
      $\fftpadForward(\xf^T[k][j],\xU^T[k][j])$\;
    }
  }
  \Return \xf\;
\caption{cconv3({\sf f},{\sf g}) 
returns an in-place implicitly dealiased convolution of
\hbox{$m_x\times m_y\times m_z$} matrices {\sf f} and {\sf g} using temporary
\hbox{$m_x\times m_y\times m_z$} matrices ${\sf U}$ and~${\sf V}$, 
$m_y\times m_z$ matrices ${\sf u}_2$ and ${\sf v}_2$,
and vectors ${\sf u}_1$ and ${\sf v}_1$ of length~$m_z$.}\label{cconv3}
\end{function}
\end{minipage}
\begin{minipage}{0.46\linewidth}
\begin{procedure}[H]
  \KwIn{vector \xf}
  \KwOut{vector \xf, vector \xu}
  $\xu[0]=\xf[0]=0$\;
  \For{$k=1$ \KwTo $2m-1$}{
    $\xu[k] \leftarrow -i\z_{4m}^k\xf[k]$\;
  }
  $\xf \leftarrow \fft\inv(\xf)$\;
  $\xu \leftarrow \fft\inv(\xu)$\;
  \Return f;
  \caption{fft0tpadBackward({\sf f},{\sf u}) stores the 
scrambled signed~$4m$-padded centered backward Fourier transform values of a
vector {\sf f} of length~$2m$ in {\sf f} and an auxiliary vector~{\sf u} of
length $2m$.}\label{fft0tpadBackward}
\end{procedure}
\begin{procedure}[H]
  \KwIn{vector \xf, vector \xu}
  \KwOut{vector \xf}
  $\xf \leftarrow \fft(\xf)$\;
  $\xu \leftarrow \fft(\xu)$\;
  \For{$k=1$ \KwTo $2m-1$}{
    $\xf[k] \leftarrow \xf[k] +i \z_{4m}^{-k}\xu[k]$\;
  }
  \Return \xf/(4m)\;
  \caption{fft0tpadForward({\sf f},{\sf u}) returns the
inverse of Procedure \fftOtpadBackward({\sf f},{\sf u}).}
\label{fft0tpadForward}
\end{procedure}
\end{minipage}
\end{figure}

\section{Implicitly dealiased ternary convolutions}\label{hyperconv}
In this section, we show that implicit padding is well suited to
dealiasing the centered ternary convolution
$$
\sum_{p=-m+1}^{m-1}\sum_{q=-m+1}^{m-1}\sum_{r=-m+1}^{m-1} f_p g_q h_r\d_{p+q+r,k},
$$
which, for example, is required to compute the time evolution of the
Casimir invariant $\int \w^4\,d\vx$ associated with the nonlinearity of
two-dimensional incompressible flow expressed in terms of the scalar
vorticity~$\w$. The basic building blocks for this problem
are again the centered Fourier transform and Hermitian convolution.

\subsection{Implicit double-dealiased centered Fourier transform}
\label{fft0bi}
Here the input data length is $2m-1$, with the Fourier origin at index
$m-1$, so one needs to pad to $N\ge 4m-3$ to prevent contamination due to
wave beating.

Implicit padding is most efficiently implemented by padding the input
vector with a single zero complex word at the beginning, to yield a vector
of length $2m$, with the Fourier origin at index $m$. We choose $m$ to be
a power of $2$ and $N=4m$, with $U_k=0$ for $k=-m$ and $k\ge m$. 

On decomposing $j=2\ell+r$, where $\ell=0,\ldots, 2m-1$ and $r\in\{0,1\}$,
we find on substituting $k'=k+m$ that
\belc
u_{2\ell +r}\no=\sum_{k=-m}^{m-1}\z_{2m}^{\ell k} \z_{4m}^{rk} U_k
=\sum_{k'=0}^{2m-1}\z_{2m}^{\ell (k'-m)} \z_{4m}^{r(k'-m)} U_{k'-m}
=(-1)^\ell i^{-r}\sum_{k=0}^{2m-1}\z_{2m}^{\ell k} \z_{4m}^{rk} U_{k-m}.
\label{fft0tbackward}
\eelc
The forward transform is then given for $k=-m+1,\ldots,m-1$ by
\bel
{4m}U_k=\sum_{r=0}^{1}\zeta_{4m}^{-rk}\sum_{\ell=0}^{2m-1}\zeta_{2m}^{-\ell k} u_{2\ell+r}
=\sum_{r=0}^{1}\zeta_{4m}^{-r(k'-m)}\sum_{\ell=0}^{2m-1}\zeta_{2m}^{-\ell (k'-m)} u_{2\ell+r}
=\sum_{r=0}^{1}\zeta_{4m}^{-rk'}i^r\sum_{\ell=0}^{2m-1}\zeta_{2m}^{-\ell k'}(-1)^\ell u_{2\ell+r},
\qquad k'\no =1,\ldots,2m-1.\label{fft0tforward}
\eel
For a ternary convolution, the product of the three factors $(-1)^\ell$
(one for each input vector) arising from \Eq{fft0tbackward}
and the factor $(-1)^\ell$ in \Eq{fft0tforward} cancels.
Procedures {\tt\ref{fft0tpadBackward}} and {\tt\ref{fft0tpadForward}}
each have an operation count asymptotic to $4Km\log_2 m$. As they
operate fully in place on their arguments, with no additional storage
requirements, it is straightforward to implement strided multivector
versions of these algorithms.

\subsection{Implicitly dealiased centered Hermitian 1D ternary convolution}
Let us now consider a centered Hermitian ternary convolution with $N=4m$,
where $m$ is a power of $2$. For explicit padding, one needs to pad the $m$
non-negative wavenumbers with $m+1$ zeros, for a total vector length of $2m+1$.

On decomposing $j=2\ell+r$, where
$\ell=0,\ldots, 2m-1$ and $r\in\{0,1\}$, the backward transform is given
by
\bec
u_{2\ell +r}=\sum_{k=-m}^{m-1} \z_{2m}^{\ell k} \z_{4m}^{rk} U_k.
\eec
If we set $U_m=0$, the real components $u_{2\ell +r}$ can thus be computed
by taking a complex-to-real transform of
$\{\z_{4m}^{rk} U_k : k=0,\ldots, m\}$.

The forward transform is
\be
{4m}U_k=\sum_{r=0}^{1}\zeta_{4m}^{-rk}
\sum_{\ell=0}^{2m-1} \zeta_{2m}^{-\ell k} u_{2\ell+r},
\qquad k\no =-m+1,\ldots,m-1.
\ee
The resulting implicitly padded centered Hermitian ternary convolution,
Function {\tt\ref{tconv}}, has an operation count of $8Km\log_2 m$. 
Five of the eight required Fourier transforms can be done out of place.
In Fig.~\ref{timing1t} we show that this algorithm is competitive with
explicit padding. Function {\tt\ref{tconv}} requires $6(m+1)$ complex words of
storage, slightly more than the $3(2m+1)=6m+3$ complex words needed for explicit
padding.

Just as for convolutions, the performance and memory benefits of
dealiasing higher-order convolutions {\it via\/} implicit padding manifest
themselves only in higher dimensions. For example, in Fig.~\ref{timing2t},
we observe for $m_x=m_y=4096$ that the implicit $(2m_x-1)\times m_y$ centered
Hermitian ternary convolution computed with Function~{\tt\ref{tconv2}} is
$2.28$ times faster than an explicit version.
The memory usage for a $(2m_x-1)\times m_y$ implicit centered Hermitian ternary
convolution is $6\cdot 2m_x(m_y+1)+3(m_y+1)=12m_xm_y+12m_x+3m_y+3$ complex words,
compared with $3\cdot 4m_x(2m_y+1)=24m_xm_y+12m_x$ complex words
(using power-of-two transforms) for an explicit version. In contrast,
the corresponding $y$-pruned convolution requires the
same amount of storage as, but is $1.42$ times faster than, an explicitly
padded version.

\begin{figure}[htbp]
\begin{minipage}{0.48\linewidth}
\begin{function}[H]
  \KwIn{vector \xf, vector \xg, vector \xh}
  \KwOut{vector \xf}
  \For{$k=0$ \KwTo $m-1$}{
    $\xu[k] \leftarrow \z_{4m}^k \xf[k]$\;
    $\xv[k] \leftarrow \z_{4m}^k \xg[k]$\;
    $\xw[k] \leftarrow \z_{4m}^k \xh[k]$\;
  }
  \medskip
  $\xu[m]=\xv[m]=\xw[m]=0$\;
  $\xu \leftarrow \crfft\inv(\xu)$\;
  $\xv \leftarrow \crfft\inv(\xv)$\;
  $\xw \leftarrow \crfft\inv(\xw)$\;
  $\xv \leftarrow \xu * \xv * \xw$\;
  $\xu \leftarrow \rcfft(\xv)$\;
  \medskip
  $\xf[m]=\xg[m]=\xh[m]=0$\;
  $\xv \leftarrow \crfft\inv(\xf)$\;
  $\xw \leftarrow \crfft\inv(\xg)$\;
  $\xg \leftarrow \crfft\inv(\xh)$\;
  $\xv \leftarrow \xv * \xw * \xg$\;
  $\xf \leftarrow \rcfft(\xv)$\;
  \medskip
  \For{$k=0$ \KwTo $m-1$}{
    $\xf[k] \leftarrow \xf[k] + \z_{4m}^{-k}\xu[k]$\;
  }
  \Return f/(4m)\;
\caption{tconv({\sf f},{\sf g},{\sf h},{\sf u},{\sf v},{\sf w}) computes
an in-place implicitly dealiased centered Hermitian ternary convolution of
three centered Hermitian vectors {\sf f}, {\sf g}, and {\sf h}, using three
temporary vectors {\sf u}, {\sf v}, and {\sf w}, each of
length~$m+1$.}\label{tconv}
\end{function}
\end{minipage}
\begin{minipage}{0.5\linewidth}
\begin{function}[H]
  \KwIn{matrix \xf, matrix \xg, matrix \xh}
  \KwOut{matrix \xf}
  \For{$j=0$ \KwTo $m_y-1$}{
    $\fftOtpadBackward(\xf^T[j],\xU^T[j])$\;
    $\fftOtpadBackward(\xg^T[j],\xV^T[j])$\;
    $\fftOtpadBackward(\xh^T[j],\xW^T[j])$\;
  }
  \For{$i=0$ \KwTo $2m_x-1$}{
    $\tconv(\xf[i],\xg[i],\xh[i],\xu,\xv,\xw)$\;
    $\tconv(\xU[i],\xV[i],\xW[i],\xu,\xv,\xw)$\;
  }
  \For{$j=0$ \KwTo $m_y-1$}{
    $\fftOtpadForward(\xf^T[j],\xU^T[j])$\;
  }
  \Return \xf\;
\caption{tconv2({\sf f},{\sf g},{\sf h}) 
returns an in-place implicitly dealiased centered Hermitian ternary convolution
of \hbox{$2m_x\times (m_y+1)$} matrices~{\sf f},~{\sf g}, and~{\sf h} using
temporary \hbox{$2m_x\times (m_y+1)$} matrices~{\sf U}, {\sf V}, and~{\sf W} and
vectors {\sf u}, {\sf v} and~{\sf w} of length~$m_y+1$.
}\label{tconv2}
\end{function}
\end{minipage}
\end{figure}

\begin{figure}[htbp]
\begin{minipage}{0.49\linewidth}
\begin{center}
\includegraphics{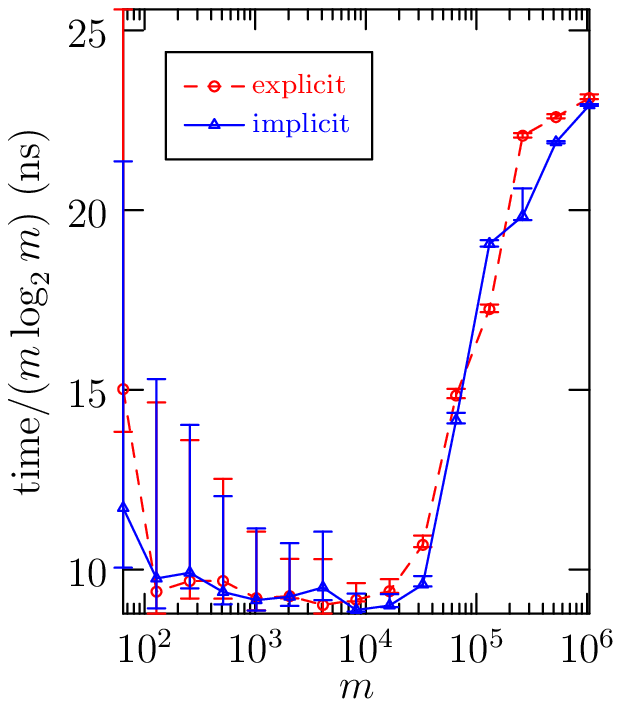}
\caption{Comparison of computation times for explicitly and implicitly
dealiased centered Hermitian in-place 1D ternary convolutions of length $m$.}
\phantomsection{}\label{timing1t}
\end{center}
\end{minipage}
\,
\begin{minipage}{0.49\linewidth}
\begin{center}
\includegraphics{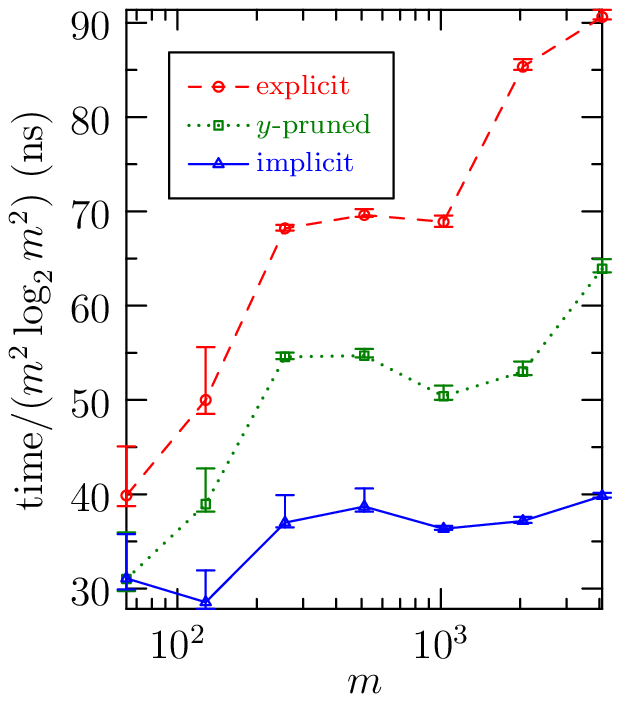}
\caption{Comparison of computation times for explicitly and implicitly
dealiased centered Hermitian in-place 2D ternary convolutions of size $(2m-1)\times m$.}
\phantomsection{}\label{timing2t}
\end{center}
\end{minipage}
\end{figure}

\section{Concluding remarks}
Explicitly padded Fourier transforms are frequently used to dealias
convolutions in one or more dimensions.
In this work we have developed an efficient method for avoiding explicit zero
padding in multidimensional convolutions, thereby saving both memory and
computation time. The key idea that was exploited was the decoupling
of temporary storage and user data, which in higher dimensions allows
the reuse of storage space. The resulting increased data locality
significantly enhanced performance by as much as a factor of $2$.
The savings in memory use, obtained by computing the Fourier transformed
data in blocks rather than all at once, was equally significant:
asymptotically, as $m_x\goesto\infty$, an implicit complex convolution
requires one-half of the memory needed for a zero-padded convolution in two
dimensions and one-quarter in three dimensions. In the centered Hermitian
case, the memory use in two dimensions is $2/3$ of the amount used for an
explicit convolution and $4/9$ of the corresponding storage requirement in
three dimensions.

Even in one dimension, where implicit padding can be implemented
competitively with conventional methods, the method has notable advantages.
For the intended application to partial differential
equations, there is flexibility in the choice of the exact convolution
size. This is why we consider for each algorithm only those vector lengths
that maximize performance.
On the other hand, for those applications where the size of the convolution
is dictated by external criteria, implicit padding effectively expands the
available set of efficient convolution sizes to include integral powers
of~$2$, a case of practical significance.
Canuto \etal~\cite[p.136]{Canuto06} point out that if only a
power-of-two transform were available for a centered convolution, zero
padding a vector of length $m=2^k$  would require a transform size of $2m$,
yielding an even slightly higher operation count, $6Km\log_2(2m)$, than the $6K
m\log_2m$ operations required for phase-shift dealiasing. The availability
of implicitly dealiased convolutions now makes this argument moot.

Another advantage of implicit padding is the ability of the algorithm to
work directly on raw unpadded user data without the inconvenience or extra
storage requirements of a separate padding buffer. 
Having a prewritten, well-tested dealiased convolution that
takes care of dealiasing internally is a major convenience for the average user.
For 2D and 3D Hermitian convolutions, a prepackaged routine should also
automatically enforce Hermitian symmetry of the data along the $x$ axis or
the~$xy$ plane, respectively. With the highly optimized implementations of
the algorithms developed in this work made available in the open source
package {\tt FFTW++}~\cite{fftwpp}, writing a pseudospectral code for
solving nonlinear partial differential equations should now be a relatively
straightforward exercise.


\bibliography{refs}
\end{document}